\documentclass[12pt,preprint]{aastex}

\lefthead{Matthews \& Karovska}
\righthead{Resolution of Mira AB at Centimeter Wavelengths}

\begin{document}

\newcommand{\ang}{\rm \AA}
\newcommand{\msun}{M$_\odot$}
\newcommand{\lsun}{L$_\odot$}
\newcommand{\days}{$d$}
\newcommand{\degree}{$^\circ$}
\newcommand{\ud}{{\rm d}}
\newcommand{\as}[2]{$#1''\,\hspace{-1.7mm}.\hspace{.0mm}#2$}
\newcommand{\am}[2]{$#1'\,\hspace{-1.7mm}.\hspace{.0mm}#2$}
\newcommand{\lsim}{~\rlap{$<$}{\lower 1.0ex\hbox{$\sim$}}}
\newcommand{\gsim}{~\rlap{$>$}{\lower 1.0ex\hbox{$\sim$}}}
\newcommand{\HII}{\mbox{H\,{\sc ii}}}
\newcommand{\kms}{\mbox{km s$^{-1}$}}
\newcommand{\HI}{\mbox{H\,{\sc i}}}

\title{First Resolved Images of the Mira~AB Symbiotic Binary at Centimeter
Wavelengths}
\author{Lynn D. Matthews\altaffilmark{1} and Margarita 
Karovska\altaffilmark{1}}

\altaffiltext{1}{Harvard-Smithsonian Center for Astrophysics,
60 Garden Street, Cambridge, MA, USA 02138}

\begin{abstract}
We report the first spatially resolved radio continuum measurements
of the Mira~AB symbiotic binary system, based on
observations obtained with the Very Large Array (VLA). This is the
first time that a symbiotic binary has been resolved unambiguously
at centimeter wavelengths.
We describe the results of VLA monitoring
of both stars over a ten month period, together with 
constraints on their individual spectral energy
distributions, variability, 
and radio emission mechanisms. The emission from Mira~A
is consistent with originating from a radio photosphere, 
while the emission from Mira~B appears best explained as free-free
emission from an ionized circumstellar region
$\sim$(1-10)$\times10^{13}$~cm in radius. 
\end{abstract}


\section{Introduction}
Symbiotics are interacting binary systems in which a cool,
evolved giant transfers material onto a hotter, compact companion
through a stellar wind (e.g., Whitelock 1987). This symbiosis
affects the late-stage evolution of both stars and their
surrounding medium and may play an important role in
shaping the formation of planetary nebulae and in the 
triggering of Type~Ia supernovae (e.g., Munari \& Renzini 1992;
Corradi et al. 2000).
Mira~AB is the nearest example of a weakly symbiotic
binary. It
comprises a pulsating asymptotic giant branch (AGB) star
(Mira~A=$o$~Ceti, the prototype of Mira variables) and a low-mass, 
accreting companion
(Mira~B, possibly a white dwarf). The pair has a projected separation of
$\sim$\as{0}{5} 
($\sim$65~AU; Karovska et al. 1997)\footnote{All physical quanitities
quoted in the this paper have been scaled to the {\it Hipparcos} 
distance of 128~pc.}, making this one
of the very few wind-accreting binaries in which the
components can be spatially and spectrally resolved with current
telescopes. This system therefore
provides a rare opportunity to study
the individual components of an interacting binary.

Multiwavelength observations of Mira AB have revealed a complex 
interacting system with tremendous temporal 
changes in the components and the circumbinary
environment (e.g., Karovska et al. 1997,2001,2005; Wood \& Karovska
2004 and references therein). 
Much of this activity is driven by mass-loss from 
Mira~A (${\dot
M}\approx2.8\times10^{-7}~M_{\odot}$ yr$^{-1}$)
through a cool,
low-velocity wind ($V_{\infty}\approx 5$~\kms; Bowers \& Knapp
1988). Material from the wind is accreted onto Mira~B, forming
a hot accretion disk, as evidenced by the presence of 
numerous rotationally broadened,
ultraviolet (UV) emission lines (Reimers \& Cassatella 1985).

{\it Chandra} observations carried out in 2003 December 
detected an
unprecedented X-ray outburst from Mira~A (Karovska et al. 2005). 
This was followed by an increase in UV emission and 
H$\alpha$ flaring lasting for about one month.
This outburst is  very unusual,
since the X-rays seem to be originating
from the AGB star rather than from the accretion disk of Mira~B.
The outburst may be associated with
a magnetic flare followed by a mass ejection event.
To constrain the origin of the outburst
and monitor its time evolution,
we subsequently began multiwavelength monitoring of the 
Mira~AB system, including 
radio continuum measurements with the 
Very Large Array (VLA)\footnote{The Very Large Array of 
the National Radio Astronomy
Observatory is a facility of the
National Science Foundation, operated under cooperative agreement by
Associated Universities, Inc.}.

\section{Observations}
We observed Mira~AB with the VLA during  ten 
epochs between 2004 October 19 and 2005 July 4 (Table~1).
Data were obtained in 4-IF continuum mode 
at four frequencies (8.5, 14.9,
22.5, and 43.3~GHz), each with a 100~MHz total bandwidth. 
Four different  array configurations were used
(A, BnA, B, and CnB), providing maximum baselines
of 36~km, 22~km, 11~km, and 6.5~km, respectively. 
In some cases, the angular resolution was insufficient 
to fully resolve Mira~AB (see Table~1). 
Our initial goal at the VLA had been monitoring Mira~A at 8.5~GHz;
however, upon the  detection of radio emission from
Mira~B, we began obtaining data at additional frequencies 
in order to constrain the spectral energy
distributions of the two sources. 
Mira~A was detected at all four observed frequencies, 
while Mira~B was detected in all but the 14.9~GHz observations. 

The data were calibrated and reduced 
using standard techniques within the Astronomical Image Processing System
(AIPS). Absolute flux levels were established through
observations and source models of 3C48 (see Table~1).
Interferometric
amplitudes and phases were calibrated by using fast switching between
Mira and the neighboring point source, 0215-023. A cycle time of
120~s was used to ``stop'' any rapid tropospheric
phase fluctuations, as emission in the
Mira field was not strong enough to permit self-calibration.
Flux densities for 0215-023
derived from each of our observations are summarized in Table~1. 
The radio images presented in this paper were computed with
the standard CLEAN deconvolution algorithm within AIPS, using a robustness
factor ${\cal R}=+1$ (Briggs 1995) to weight the
visibilities. 

\section{Analysis and Discussion}
Figure~1 shows several examples of our recently obtained 
radio continuum images of Mira~AB. 
Two continuum sources are clearly detected in each image. 
After accounting for proper motion, the two sources correspond with 
the positions of Mira A and B, respectively, as recently determined
from {\it HST} UV observations (Figure~2).
This is the first time the components of a symbiotic
binary have been resolved unambiguously at centimeter
wavelengths.

For each observation, we measured the flux densities of the two
stellar components by fitting elliptical Gaussians to the data in the
image plane. Our results are summarized in Table~1. 
Figure~3 plots the derived flux densities of Mira~A and B as a
function of time. 

With one exception, 
we found Mira~A to be brighter than Mira~B at all times and at
all frequencies. However,
our 8.5~GHz measurements on JD~2,453,394 seem to indicate
a simultaneous brightening of Mira~B and dimming of Mira~A compared with our
previous measurements,  while nine days later, the 
flux densities of the two are nearly identical.   
We believe these apparent synchronous variations to be an artifact of 
undersampling the $u$-$v$ plane, and 
were able to reproduce an analogous ``redistribution'' 
in flux using pairs of similarly separated artificial point
sources introduced into the visibility data. Note however that 
total flux density (A+B) was conserved. The observations in question
were obtained using the BnA configuration of the VLA, which
produced a highly elongated beam pattern, with the
point spread functions for Mira~A \& B partially overlapping at 8.5~GHz.

Our observations spanned 
$\sim78$\% of the 332 day pulsation period of Mira~A, during which the
optical brightness of the star varies by $\sim$6-7 magnitudes 
(e.g., Reid \& Goldston 2002). 
Our measurements imply that any changes in the radio brightness of
either star linked to these pulsations have a
substantially smaller amplitude. At
8.5~GHz, where we have the longest observational baseline, the
data are consistent with brightness changes of $\lsim$30\% from the mean
for both Mira~A \& B during the 189-day interval between 2004 October 19
and 2005 April 26 ($\lsim$15\% for Mira~A excluding the BnA array data;
see above). 

We have used a weighted mean of the measurements at each frequency  
where the  components of Mira~AB were resolved  to
estimate the spectral indices of the two stars. 
Figure~4 shows the resulting radio frequency spectra.
Both Mira~A and B have positive spectral indices,
$\alpha$ (where  flux density
$S_{\nu}\propto \nu^{\alpha}$), indicating that the emission
mechanism for both sources is likely to be predominantly thermal.
Nonlinear least-squares fits to the data show the
radio spectra of Mira~A and B can be described 
as $S_{\nu,A}=(0.009\pm0.004)\nu^{1.50\pm 0.12}_{\rm GHz}$~mJy and
$S_{\nu,B}=(0.010\pm0.008)\nu^{1.18\pm0.28}_{\rm GHz}$~mJy, respectively.

The initial goal of our radio observations of Mira~AB was to
search for radio signatures of the recent X-ray outburst detected by
Karovska et al. (2005). From the continuum observations obtained to date, 
we have not detected any unambiguous
aftereffects of this event. 
Some of our radio images of Mira~A (e.g., Figure~1) show 
elongations along a position angle of $\sim120^{\circ}$,
analogous to those seen in X-rays and in the
UV (Karovska et al. 2005). These features could be consistent with an
outflow; however, given the phase noise and limited $u$-$v$ sampling 
of our data (observations 
$\lsim$1 hour), they may also be spurious.
If the X-ray outburst of Mira~A was connected with 
a mass ejection event and/or shock formation, we
might also expect a brightening in one or both components of
Mira~AB. We detected some signs of statistically signficant flux density
fluctuations in Mira~A and in the total flux from the binary at
8.5, 22.5, and 43.3~GHz; however,
these are generally limited to a single frequency on a given date, and cannot be
unambiguously linked with the outburst. Future monitoring
should reveal whether similar-scale fluctuations  are the norm
for the system. 
We note that  material
ejected from Mira~A at speeds of $\sim$200~\kms\ 
(see Karovska et al. 2005) would require $\gsim$1.5 years to reach
the accretion disk of 
Mira~B. Therefore radio signatures of such an event may still occur
during late 2005 or 2006. 

To gauge the longer-term variability of the Mira~AB system,
we have also analyzed unpublished archival 8.5~GHz data 
taken in 1996 December using the VLA in its A
configuration. Both
Mira~A and Mira~B were detected with
flux densities $S_{A}=0.28\pm 0.05$~mJy and $S_{B}=0.15\pm
0.04$~mJy, respectively. These values are consistent with 
our recent measurements, and imply changes in the mean brightness 
of $\lsim$30\% for Mira~A and $\lsim$20\% for Mira~B during
the past 8.5 years. 

Previous VLA 8.4~GHz measurements of Mira~A have been published by
Reid \& Menten (1997), based on data obtained during 1990. However,
Mira A \& B were
spatially unresolved in these observations, thus the authors likely 
measured the {\it combined} emission from the two binary components. Indeed,
their published flux densities agree  with the values we derive
from the sum of Mira~A+B. 
Reid \& Menten (1997) interpreted the radio emission they observed
from Mira and other long period variable stars as arising from radio
photospheres located at $R\approx2R_{\star}$. 
Their models predict that at centimeter
wavelengths ($\sim$8-22~GHz) this emission   
should exhibit a spectral index $\alpha\approx$1.86.
This is slightly steeper than the  value we derive for Mira~A
(Figure~4). However,
after accounting for calibration uncertainties, it is
consistent with our data in the range 8.5-22.5~GHz. 
At higher frequencies, a gradual turnover may be expected
owing to opacity changes, consistent with our 43.3~GHz measurements.
The Reid \& Menten model also
predicts an absolute 8.5~GHz flux density for Mira~A in agreement with
our mean observed value.
Our new unblended radio continuum measurements of Mira~A therefore remain 
consistent with 
the emission  arising primarily from a radio
photosphere. The limits on brightness fluctuations of $\lsim$30\% over the
course of several months furthermore place a limit on shocks or other
disturbances in the radio photosphere of Mira~A (e.g., resulting from pulsations
or the X-ray outburst event)  to speeds of less than a few
\kms\ (see Reid \& Menten 1997). 

In the case of Mira~B, thermal emission from the stellar photosphere
and/or accreting surface will be undetectable at centimeter wavelengths
owing to the small emitting area. However, Lyman continuum photons 
from its hot accretion disk or boundary layer may 
ionize a portion of Mira~A's wind, providing a source of
free-free emission. In general, free-free radiation from 
a fully ionized stellar wind results in  radio emission with a
spectral index $\alpha=0.6$
(Wright \& Barlow 1975). Our measured spectral index for Mira~B is a
factor of two
steeper than this, and is also steeper than values previously
measured for other (unresolved) symbiotics over this frequency range 
(Seaquist \&
Taylor 1990). However, in the limit where only a very small fraction
of the wind is ionized, the model of Seaquist et al. (1984) and 
Taylor \& Seaquist (1984) (the so-called ``STB'' model) predicts
$\alpha\rightarrow$1.3-2.0 across the optically thick portion of the
spectrum, consistent with our measurements. 
Assuming a canonical electron temperature of
$T_{e}=10^{4}$~K (Osterbrock 1989), the  radius of 
an optically thick ionized sphere required to produce the observed
$\nu$=8.5~GHz emission from Mira~B 
can be estimated as $R=[S_{\nu}c^{2}d^{2}/2kT_{e}\nu^{2}\pi]^{0.5}\approx
1.6\times10^{13}$~cm ($\sim$1~AU) where $c$ is the speed of light and 
$k$ is Boltzmann's constant. This suggests an
ionized volume much smaller than the binary separation.

An alternate constraint on the size of an ionized region surrounding
Mira~B may be obtained by incorporating results from previous studies.
Based on UV spectroscopy, Reimers \& Cassatella (1985)
estimated a radius and thickness 
for the Mira~B accretion disk of $r\sim1.7\times10^{11}$~cm and
$t\sim6.8\times10^{9}$~cm, respectively,
and an electron temperature $T_{e}=11,000$~K. A blackbody with this
emitting area and temperature will produce
$N_{\rm UV}\approx5\times10^{41}$ hydrogen-ionizing photons per
second. 
Assuming a separation between Mira~A \& B of $a$=100~AU (slightly larger than
the projected separation), we can also estimate the particle density from
the wind of Mira~A at the location of Mira~B as
$n_{e}={\dot M}/4\pi\mu m_{\rm
H}V_{\infty}a^{2}$ (Wright \& Barlow 1975, Eq.~2), where ${\dot M}$ and
$V_{\infty}$ are Mira~A's wind parameters, $\mu$ is the mean molecular
weight ($\sim$1), and $m_{H}$ is the mass of a hydrogen atom. 
This predicts $n_{e}\approx7.2\times10^{5}$~cm$^{-3}$, neglecting
density enhancements caused by the gravitational field of 
Mira~B. In 
the idealized case of a pure hydrogen medium of uniform
density, 
the size of the resulting ionized region as predicted by
the Str\"omgren relation (e.g., Osterbrock 1989) is $R_{s}\approx(3N_{\rm
UV}/4\pi\alpha_{r}n^{2}_{e})^{\frac{1}{3}}\approx9.8\times10^{13}$~cm. 
Here $\alpha_{r}$ is the
recombination coefficient to  all levels but the ground state of
hydrogen ($\approx2.6\times10^{-13}$~cm$^{3}$~s$^{-1}$; Osterbrock 1989).
Although this order-of-magnitude 
estimate neglects such factors as a non-uniform density within
the ionized volume and the presence of molecules and heavier
elements in the wind, it  supports the STB model as an explanation for the radio
emission from Mira~B and
reaffirms that any ionized nebula around the star should be 
quite small compared with the binary separation and the overall
extent of Mira~A's circumstellar envelope ($>10^{17}$~cm;
Bowers \& Knapp 1988). It is also consistent with the nebula being
unresolved by  our recent VLA observations (predicted angular diameter 
$\theta\lsim$\as{0}{1}) . Future VLA
measurements should provide additional constraints on the
size and shape of this ionized region, and ultimately place new,
independent constraints on the Mira~B accretion disk properties and the
true binary separation.

We have considered other possible mechanisms as the source of the
radio emission from Mira~B, but none appear likely to contribute
appreciably to the observed flux. 
For example, if Mira~B is a magnetically active dwarf rather than a
white dwarf (Jura \& Helfand 1984; Kastner \&
Soker 2004), its
quiescent radio luminosity is expected be $\sim10^{12}$-$10^{14}$ erg
s$^{-1}$ Hz$^{-1}$ (Benz \& G\"udel 1994)---undetectable at Mira~B's distance.
Mira~B is also known to power a variable 
wind  (Wood et al. 2002), but even if fully ionized, it is too feeble
to produce detectable radio emission. Based on the parameters derived
by Wood et al. (2002):
${\dot M}\approx$(0.14-2.8)$\times10^{-11}~M_{\odot}$ 
yr$^{-1}$ and $V_{\infty}$=250-400~\kms, the predicted 
radio flux density
from Mira~B's wind (see Wright \& Barlow 1975, Eq.~8) 
would be $\sim$4 orders of magnitude smaller than the
observed emission. 
Finally, the steep spectral index  of the radio emission from
Mira~B ($\alpha>$0.6) and low energy flux of its wind 
argue against any significant contribution from non-thermal
emission produced by collisions between the two winds of the 
binary (see Dougherty \& Williams 2000;
Kenny \& Taylor 2005). 

In summary, we have presented new centimeter wavelength images of 
the symbiotic binary system,  Mira~AB. These data allow for the first time
measurements of the
radio properties of the individual components of a symbiotic binary.
Over the frequency range 8-43~GHz, the radio properties of
the evolved giant,
Mira~A, are consistent with the radio emission originating
predominantly from a radio
photosphere. The emission from Mira~B is consistent with
arising from a circumstellar 
\HII\ region $\sim$(1-10)$\times10^{13}$~cm in radius. We find the
radio variability of both stars to be $\lsim$30\% over a ten month
period during 2004-2005. Flux densities we derived from 
archival data taken in 1996 are consistent with
our recent measurements, and imply changes in the mean 8.5~GHz flux densities of
$\lsim$30\% for Mira~A  and $\lsim$20\% for Mira~B during the past 8.5
years.

\acknowledgements
We thank M. Reid for valuable discussions, and are grateful to the VLA
staff for their support of this project.
LDM was funded by a Clay Fellowship from the Smithsonian
Astrophysical Observatory (SAO).
MK is a member of the {\it Chandra} X-ray Center, which is operated
by SAO under contract to NASA NAS8-39073.

\clearpage

\begin{figure}
\scalebox{0.60}{\rotatebox{-90}{\includegraphics{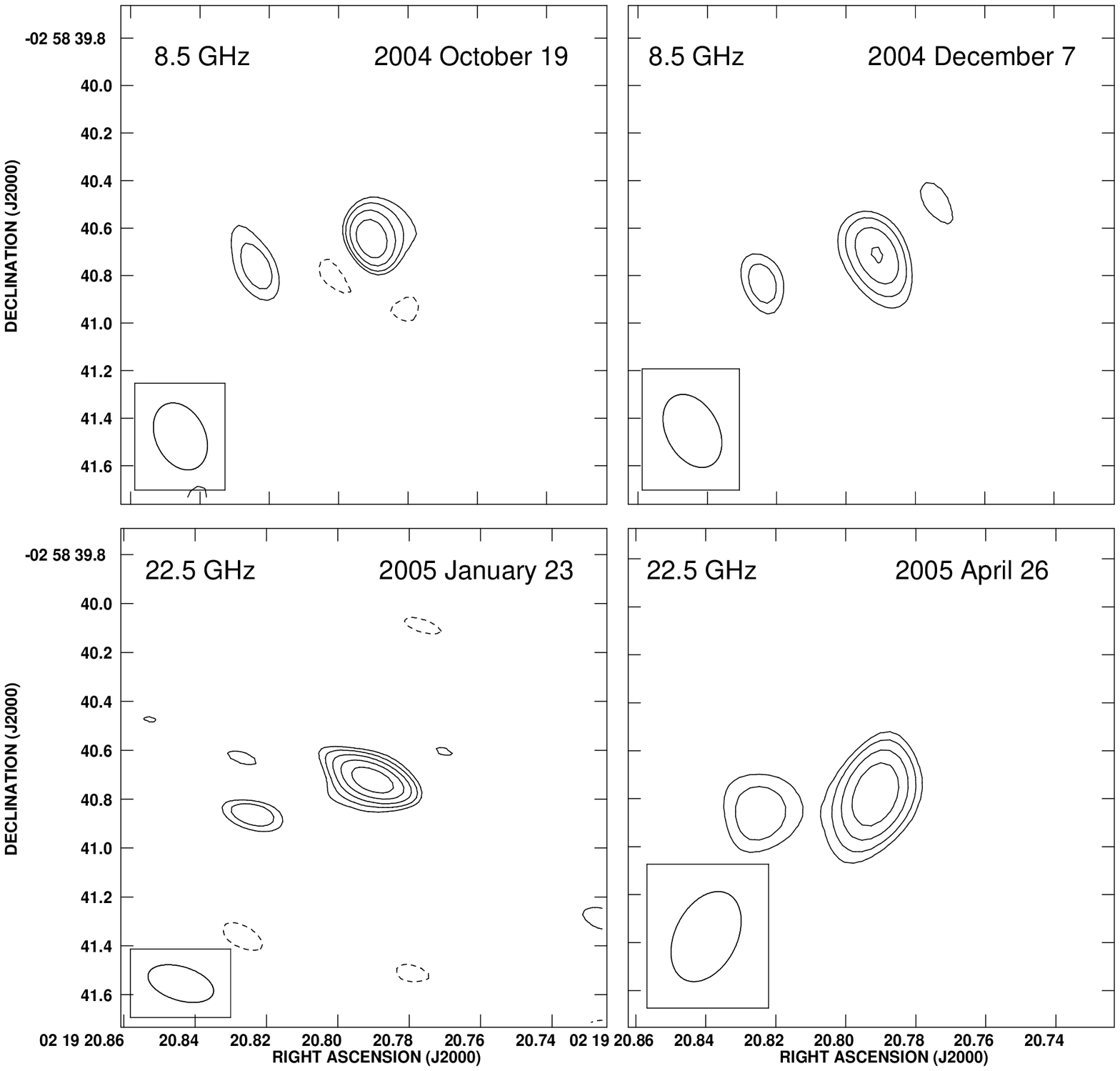}}}
\caption{Radio continuum images of Mira~AB obtained with the VLA at
8.5~GHz (top) and 22.5~GHz (bottom) on four different 
dates. Mira~A is near the center of each panel, and Mira~B is $\sim$\as{0}{5}
to the southeast. Contour levels are
(-4.2,-3.0,3.0,4.2,6,8.5)$\times$0.025 mJy beam$^{-1}$  for the 8.5~GHz
images and (-4.2,-3.0,3.0,4.2,6,8.5,12)$\times$0.066 mJy beam$^{-1}$ for
the 22.5~GHz images.  The lowest contour levels are $\sim3\sigma$. The
restoring beams are shown in the lower left corner of each panel. Both stars
are consistent with point sources.}
\label{mira_allfreq}
\end{figure}

\clearpage

\begin{figure}
\centering
\scalebox{0.7}{\rotatebox{0}{\includegraphics{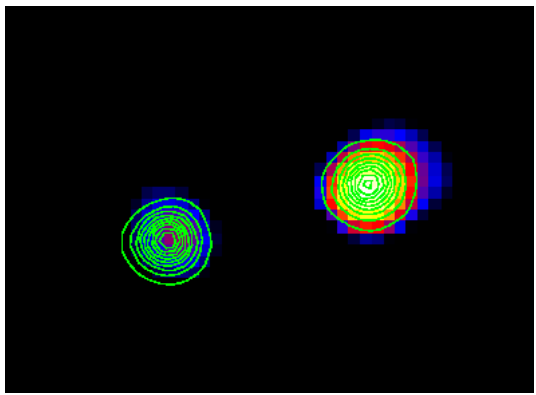}}}
\caption{Contours of an {\it HST}
3729\ang\ image of Mira~AB from 2004 February (Karovska et al. 2005) 
overlaid on a VLA 8.5~GHz 
image of Mira~AB from 2004 October. The VLA image was restored using 
a circular beam with FWHM \as{0}{25}. The {\it HST} 
contours show 10\% brightness 
increments. For display purposes, the intensity peaks of Mira~A 
have been aligned in the two images. The field-of-view is
$\sim$\as{1}{2}$\times$\as{0}{85}.
North is on top, east to the left.  }
\label{mira_8ghz+cha}
\end{figure}

\clearpage

\begin{figure}
\scalebox{0.75}{\rotatebox{90}{\includegraphics{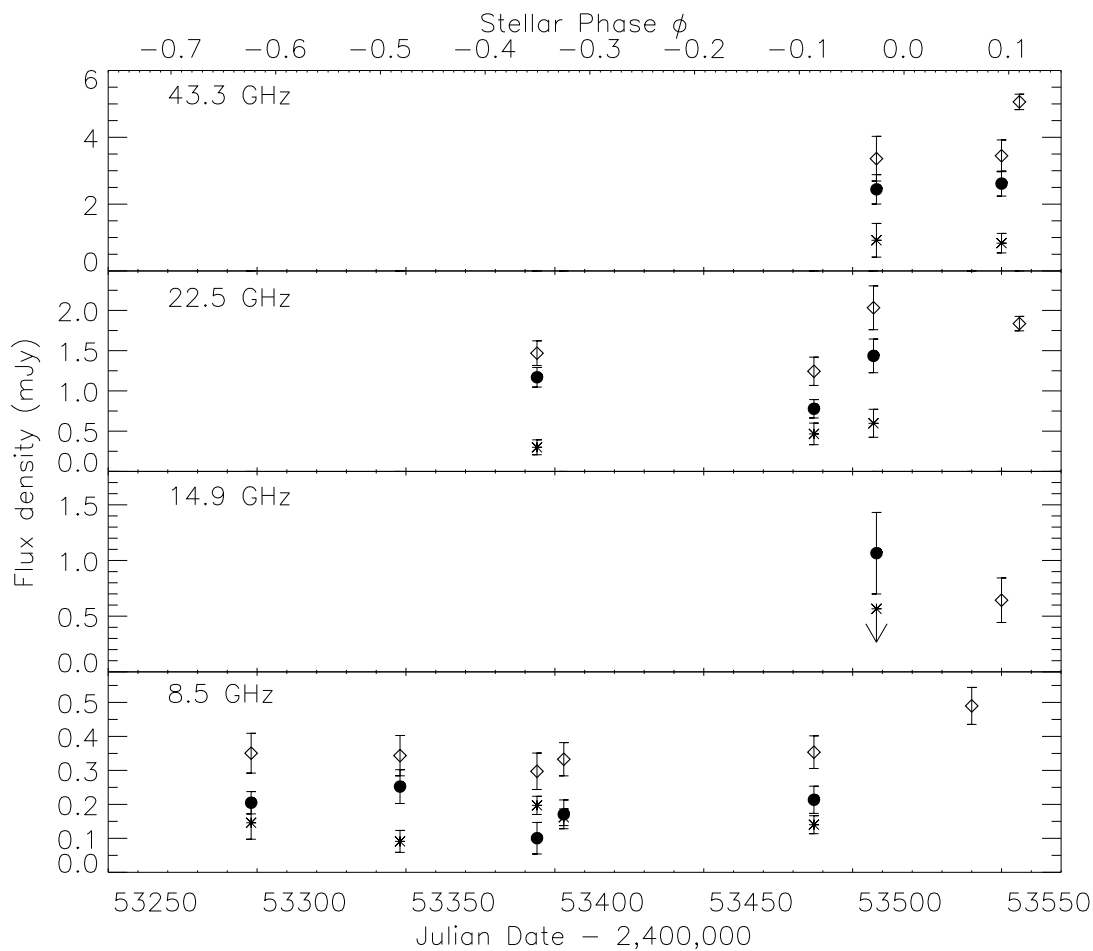}}}
\caption{
Plots of radio continuum flux density in mJy 
as a function of time for the two
components of Mira~AB  at four frequencies. Values for 
Mira~A are plotted as closed circles,
and those for Mira~B as asterisks. Diamonds represent the sum of the 
two components. In cases where A \& B were unresolved,
only the combined flux density is shown. The 1$\sigma$ error 
bars indicate statistical uncertainties but
do not include calibration uncertainties. The optical stellar phase of
Mira~A is shown along the upper axis; its optical maximum 
occurs at $\phi$=0, and  optical minimum 
at $\phi=-$0.36.
The individual 8.5~GHz flux densities for the two stellar
components on 
JD~2,453,394 and 2,453,403 are unreliable (see Text). }
\label{mira_allfreq}
\end{figure}

\clearpage

\begin{figure}
\scalebox{0.65}{\rotatebox{90}{\includegraphics{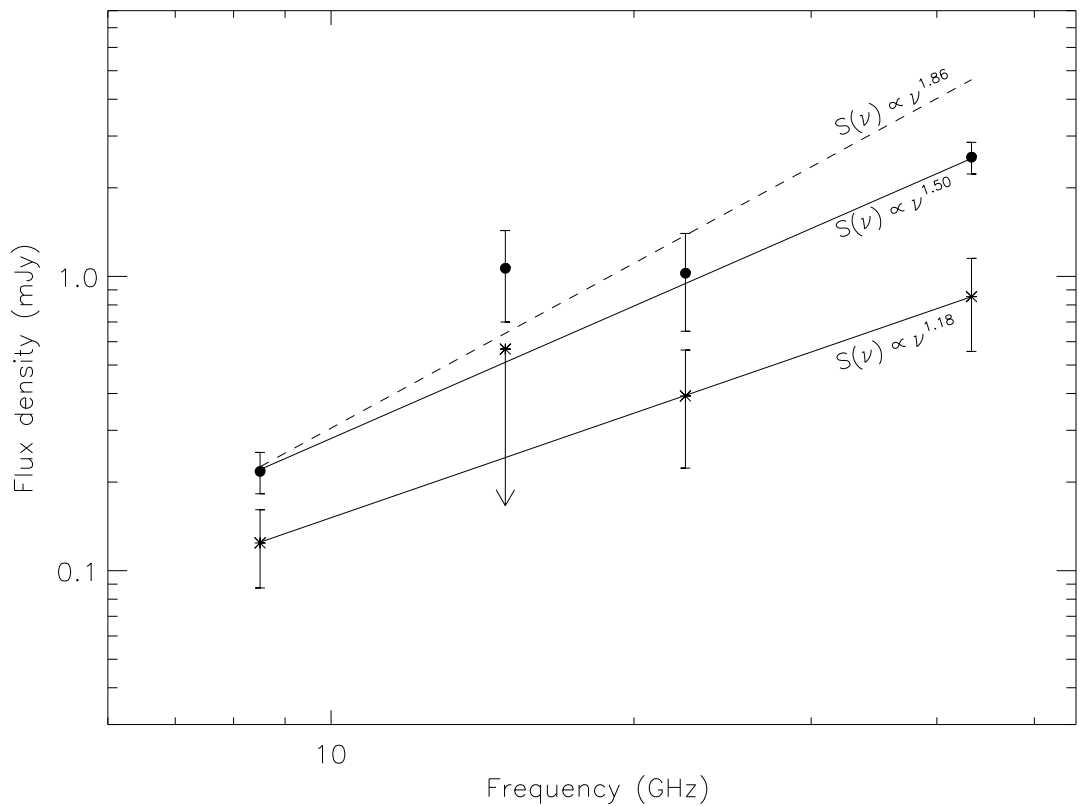}}}
\caption{Radio frequency spectra for Mira~A (closed circles) and
Mira~B (asterisks). Points at each frequency represent a weighted mean of all
data from Table~1 for which the stellar components could be
resolved sufficiently  to permit individual flux density
measurements. Error bars include both measurement scatter and
systematic uncertainties.
A 3$\sigma$ upper limit is shown for Mira~B at 14.9~GHz.
Overplotted are the best-fitting power laws of the form
$S(\nu)\propto\nu^{\alpha}$ (solid lines), as well as the predicted spectrum
based on the static 
photosphere model of Reid \& Menten 1997 (dashed line). }
\label{mira_flux+fit}
\end{figure}

\clearpage

%
\begin{deluxetable}{lcrcccccc}
\tabletypesize{\scriptsize}
\tablewidth{0pc}
\tablenum{1}
\tablecaption{VLA Monitoring Data for Mira A \& B}
\tablehead{
\colhead{Julian} & \colhead{Array} & \colhead{$\nu$} &
\colhead{Image rms} &
\colhead{Flux Density} & \colhead{Flux
Density} & \colhead{Flux Density}
& \colhead{Flux Density$^{a}$}  & \colhead{Resolved?}\\
\colhead{Date}      & \colhead{Configuration}  & \colhead{(GHz)} &
\colhead{(mJy beam$^{-1}$)} &
\colhead{A (mJy)} & \colhead{B (mJy)} & \colhead{A+B (mJy)} &
\colhead{0215-023 (Jy)}  & \colhead{} }

\startdata

2,453,298 & A & 8.5 & 0.03 & 0.20$\pm$0.03 & 0.14$\pm$0.05 & 0.34$\pm$0.06 &
0.765$\pm$0.004 & F \\

2,453,348 & A & 8.5 &0.02  & 0.25$\pm$0.05 & 0.09$\pm$0.03 & 0.34$\pm$0.06 &
0.764$\pm$0.005 & F \\

2,453,394 & BnA & 8.5 & 0.03 & 0.10$\pm$0.05$^{b}$ & 0.20$\pm$0.03$^{b}$ &
0.30$\pm$0.05 & 0.769$\pm$0.003 & P \\

2,453,394 & BnA & 22.5 &  0.07 & 1.17$\pm$0.12 & 0.30$\pm$0.09 & 1.47$\pm$0.15 &
0.500$\pm$0.005 & F \\

2,453,403 & BnA & 8.5 &0.02 & 0.17$\pm$0.04$^{b}$ & 0.16$\pm$0.02$^{b}$
&0.33$\pm$0.05& 0.794$\pm$0.002 &
P \\

2,453,487 & B & 8.5 & 0.03 & 0.21$\pm$0.04 & 0.14$\pm$0.03 & 0.35$\pm$0.05
&0.786$\pm$0.003 &  P \\

2,453,487 & B & 22.5 &0.06 & 0.78$\pm$0.11 & 0.46$\pm$0.13 & 1.24$\pm$0.18
&0.468$\pm$0.006 &
F\\

2,453,507 & B & 22.5 &0.12  & 1.44$\pm$0.21 & 0.60$\pm$0.17 & 2.04$\pm$0.27 &
0.495$\pm$0.003 &
F \\

2,453,508 & B &14.9 &0.16 & 1.06$\pm$0.37 & $<0.57^{c}$ & $<$1.63 &
0.652$\pm$0.002 & F \\

2,453,508 & B & 43.3 & 0.26 & 2.44$\pm$0.44 & 0.92$\pm$0.50 & 3.36$\pm$0.67
&0.284$\pm$0.006 & F \\

2,453,540 & CnB & 8.5 & 0.03 & ... & ... & 0.49$\pm$0.05 & 0.800$\pm$0.001 & U\\

2,453,550 & CnB & 14.9 &0.14  & ... & ... & 0.64$\pm$0.20 & 0.651$\pm$0.001 &
U \\

2,453,550 & CnB & 43.3 &0.19  & 2.61$\pm$0.37 &0.83$\pm$0.29  & 
3.44$\pm$0.47  & 0.265$\pm$0.004 &
F \\

2,453,556 & CnB & 22.5 &0.05  & ... & ... & 1.84$\pm$0.09 &0.485$\pm$0.004 & U\\

2,453,556 & CnB & 43.3 &0.13 & ... & ... & 5.06$\pm$0.23 & 0.265$\pm$0.003 &
P \\

\enddata

\tablecomments{{Julian date (JD) 2,453,298 corresponds to 2004 October 19 and
JD~2,453,556 to 2005 July 4.
The last column indicates the
degree of spatial resolution of
Mira A \& B: F=fully resolved, P=partially resolved, U=unresolved.}}
\tablenotetext{a}{Amplitude and phase
calibrator; the absolute flux scale was established using 
observations of 3C48 plus clean-component models of this source. Adopted fluxes
for 3C48 were: 3.15~Jy (8~GHz); 1.74~Jy (15~GHz), 1.12~Jy
(22~GHz), and 0.53~Jy
(43~GHz). Calibration uncertainties 
are $\sim$10\% at 8~GHz, 15~GHz, \& 22~GHz, and $\sim$20\% at
43~GHz.}
\tablenotetext{b}{Flux 
densities of individual components are unreliable (see Text).}
\tablenotetext{c}{3$\sigma$ upper limit.}

\end{deluxetable}

\end{document}